\documentclass[conference]{IEEEtran}
\IEEEoverridecommandlockouts
\usepackage{cite}
\usepackage{amsmath,amssymb,amsfonts}
\usepackage{algorithmic}
\usepackage{graphicx}
\usepackage{textcomp}
\usepackage{xcolor}
\usepackage{amsmath}
\usepackage{amsthm}
\usepackage[english]{babel}
\usepackage{xspace}
\usepackage{cases}
\usepackage{xcolor}
\usepackage{graphicx}
\usepackage[font=footnotesize]{caption}
\usepackage[font=footnotesize]{subcaption}
\usepackage{epstopdf}
\usepackage{upgreek}
\usepackage{tikz}
\usetikzlibrary{shapes.geometric, arrows}
\usepackage{booktabs}
\usepackage{array}
\usepackage{blindtext}
\usepackage{multirow}
\usepackage{bbold}
\usepackage{soul}
\usepackage{enumerate}
\usepackage{blindtext}
\usepackage{graphicx}
\usepackage{xcolor}
\usepackage{siunitx}
\usepackage{pgf}
\usepackage{lipsum}
\usepackage{glossaries}
\usepackage{enumitem}
\usepackage{setspace}
\usepackage{hyperref}

\setitemize[1]{itemsep=4.5pt,partopsep=1pt,parsep=\parskip,topsep=4.5pt}
\usepackage{algorithm,algorithmic}

\setkeys{glslink}{hyper=false}

\newacronym{opf}{OPF}{optimal power flow}
\newacronym{qp}{\textsc{qp}}{quadratic program}
\newacronym{nlp}{\textsc{nlp}}{nonlinear programming}
\newacronym{rapidpf}{rapid\textsc{pf}}{rapid prototyping for distributed Power Flow}
\newacronym{admm}{\textsc{admm}}{Alternating Direction Method of Multipliers}
\newacronym{aladin}{\textsc{aladin}}{Augmented Lagrangian based Alternating Direction Inexact Newton method}
\newacronym{ocd}{\textsc{ocd}}{Optimality Condition Decomposition}
\newacronym{app}{\textsc{app}}{Auxiliary Problem Principle}
\newacronym{sqp}{\textsc{sqp}}{Sequential Quadratic Programming}

\definecolor{RB}{rgb}{.1,.4,.9}
\definecolor{VH}{rgb}{0.,.8,.4}
\definecolor{revised}{rgb}{.2,.8,.1}

\newcommand{\matlab}{\textsc{matlab}\xspace}

\newcommand{\matpower}{\textsc{matpower}\xspace}

\def\BibTeX{{\rm B\kern-.05em{\sc i\kern-.025em b}\kern-.08em
    T\kern-.1667em\lower.7ex\hbox{E}\kern-.125emX}}
\begin{document}

\title{Industrial Application of the Shapley value-based Redispatch Cost Allocation to Large-Scale Power Grids requires AC Optimal Power Flow
}

\author{\IEEEauthorblockN{1\textsuperscript{st} Rebecca Bauer}
\IEEEauthorblockA{\textit{Institute of Automation} \\
\textit{and Applied Informatics} \\
\textit{Karlsruhe Institute of Technology}\\
Karlsruhe, Germany \\
rebecca.bauer@kit.edu}
\and
\IEEEauthorblockN{2\textsuperscript{nd} Xinliang Dai}
\IEEEauthorblockA{\textit{Institute of Automation} \\
\textit{and Applied Informatics} \\
\textit{Karlsruhe Institute of Technology}\\
Karlsruhe, Germany \\
xinliang.dai@kit.edu}
\and
\IEEEauthorblockN{3\textsuperscript{rd} Veit Hagenmeyer}
\IEEEauthorblockA{\textit{Institute of Automation} \\
\textit{and Applied Informatics} \\
\textit{Karlsruhe Institute of Technology}\\
Karlsruhe, Germany \\
veit.hagenmeyer@kit.edu}
}


\maketitle

\begin{abstract}


A burgeoning topic in the current energy transition are the huge costs of redispatch congestion management (CM) in large transmission systems. One of the German transmission system operators (TSOs) raised the critical inquiry of how to allocate the redispatch costs amongst TSOs in an equitable and beneficial way. Previously, a Shapley value-based approach has been introduced on small test grids, using the linear DC approximation of optimal power flow (OPF). However, within the application of CM, its feasibility and accuracy for large-scale power grids and its impact on the computed congestions remain uncertain. Therefore, this study investigates the applicability of the DC OPF compared to the exact AC OPF with regard to the Shapley values, for both small and large-scale grids. Numerical simulation shows significant differences in the congested lines, the overall redispatch costs, and the Shapley values. These findings suggest that for future CM, the TSOs should further investigate AC OPF solutions.

\end{abstract}


\section{Introduction}
\label{sec:introduction}

The energy transition requires more distributed renewable energy sources (RES) to be integrated into the power grids, e.g. the German grids, as well as the whole European grid.
However, their decentrality and frequent high power injections cause large power flows in the transmission grids, whose structure is still built for unidirectional power flows. With the new bidirectional power flows, power lines are regularly charged close to their capacity limits, requiring a large amount of congestion management (CM) for the transmission system operators (TSOs).
Congestion management mainly encompasses redispatch measures that shift generation from one end of the stressed lines to the other.
Redispatch costs in Germany have risen to several billion euros \cite{bnetza_bericht}.
 
There are two challenges: Firstly, costs are not fairly allocated, as TSOs who perform redispatch pay instead of those who cause the congestion. Hence, a mechanism incentivizing long-term grid expansion including fairness is needed.
Secondly, all redispatch optimal power flow (OPF) computations use a linear DC OPF approximation instead of the exact AC OPF formulation, as seen in most simulation software \cite{redispatch, NEMO}. This might lead to using more expensive control energy \cite{Diskussionspapier}. Also, contrary to the US, AC OPF is not yet pushed by the German government or the EU. 
The open question answered in the present paper is how much the DC approximation affects the redispatch costs and makes cost allocation less exact. 

Concerning the cost sharing mechanism, one idea is a market-based solution such as locational marginal prices (LMPs) \cite{LMPs} that are already used in the US. LMPs are very simple, node-based, i.e. consumer-focused, they are integrated into the real-time (RT) market for short-term CM and lead to a very efficient utilization of the power grid. However, they do not include fairness conditions and do not incentivize long-term grid expansion. 

For Germany and Europe there are further requirements: The regions have several TSO zones, cost allocation has to be ex post and independent of the markets, the approach should include a notion of fairness, allocate costs to grid operators instead of consumers, and incentivize grid expansion.
While LMPs are considered for a future European electricity market \cite{CMEurope, LMPs, EUReformElectricityMarket}, they are not a well applicable option at present \cite{NodalPricingEurope}.  

The Shapley value is a cost or profit key from the field of game theory \cite{HandbookShapley}, fulfilling several fairness axioms. 
Recent work has realized this idea for power grids in the setting of DC OPF \cite{voswinkel} by modelling the congestions in redispatch as players in a collaborative game.
Subsequently, 
\cite{bauer}
extended the formulation to AC OPF, computing the OPF in distributed fashion to respect the data privacy between the TSOs.
Other applications of the Shapley value in energy systems explore cost sharing for market flexibilities between TSO and DSO \cite{flexibility}. 
For a similar grid setting, \cite{TSODSO} compares several cost allocation methods, declaring the Shapley value as the favorable choice.
However, all previous works have in common that they focus on small test grids using the DC approximation of OPF.

Hence, in the present paper, we pose and answer the following research questions: 
\begin{itemize}
    \item[(a)] \textit{What effect has the AC OPF instead of the DC approximation on the computed congestions and the resulting Shapley values?}
    \item[(b)] \textit{How does the effect of AC OPF show on large-scale meshed transmission grids?}
\end{itemize}
Studying these questions yields counter-intuitive answers in the sense that AC OPF is not simply an extrapolation of the linear approximation, but yields very different results.

The literature sheds some light on the differences between DC and AC OPF. 
A well-known fact is that DC OPF solutions are never AC feasible \cite{DCACfeasible}, hinting that the effect of the OPF does not produce correct Shapley values. 
Further, \cite{comparison} inspects the differences in DC and AC OPF and examines AC convergence.
Finally, \cite{feasibilityOPF} states more precisely that the feasible regions of DC and AC OPF never intersect, and that decisions based on DC OPF should be taken with care, also when using small-scale grids.
Given that all of these works suggest potentially very large differences in the OPF results using the AC instead of the DC formulation, we assume these differences might also largely affect the redispatch cost allocation. 
Moreover, as most studies were performed on small-scale grids, the question of how the different formulations affect the Shapley values on large-scale grids, remains open.

The contributions of this paper are the following:
\textit{
\begin{itemize}
    \item[(1)] We show that DC OPF, compared to AC OPF, can yield a completely different congestion situation, as well as overall costs and Shapley values, and is, thereby, not accurate enough for congestion management.
    \item[(2)] We show the feasibility of the Shapley algorithm with DC and AC OPF on large-scale grids.
    \item[(3)] We present exemplary counter-intuitive effects on congestions, costs, and Shapley values arising through AC OPF.
\end{itemize}
}
The findings show the TSOs and other grid operators that for practical congestion management with the Shapley value-based approach, and potentially other cost sharing values, for a fair cost allocation, AC OPF solutions are required.

The rest of this paper is structured as follows. In Section \ref{sec:methodology} the Shapley value-based approach using the AC OPF formulation is introduced. Section \ref{sec:results} gives the numerical results and further insights on the algorithm. Section \ref{sec:conclusion} concludes this paper and gives an outlook on future work.
\section{Methodology}
\label{sec:methodology}

\subsection{Notation}
Given a grid $\mathcal{G}=(\mathcal{N},\mathcal{L})$, where $\mathcal{N}$ is the set of $N$ buses and $\mathcal{L}$ is the set of $L$ branches.
For the Shapley value we define a cooperative game $\mathcal{G}=(\mathcal{L}^{p},\,\Phi)$, following~\cite{voswinkel}. $\mathcal{L}^{p}\subseteq\mathcal{L}$ denotes the set of players $\xi:=(i,j)\in\mathcal{L}^{p}$, i.e. the set of all congested lines connecting bus $i$ and bus $j$. Coalitions, i.e. groups of congested lines, are denoted by $\Omega\in\mathbb{P}(\mathcal{L}^{p})$. $\Phi$ is the characteristic function that assigns each coalition $\Omega$ the system cost $\Phi(\Omega)$.

\subsection{Shapley Value for Cost Allocation}
The Shapley value is a unique allocation rule from the field of game theory satisfying the four fairness axioms \cite{HandbookShapley}. 
It gives a fair distribution of the total costs $\Phi(\mathcal{L}^{p})$ of the grand coalition $\mathcal{L}^{p}$. In a game $\mathcal{G}$, the Shapley value of each player $\xi$ is the expected marginal contribution of a player over all coalitions $\Omega$. 
In this paper, the total costs are the total redispatch costs, the players are the congested lines, and the Shapley value is the cost that each congested line produces.

The Shapley value of a player $\xi$ is as follows:
\begin{equation}\label{eq::shapley}
    \small
    \Psi_\xi(\Phi) = \sum_{\Omega\in\mathcal{L}^{p}\setminus \{\xi\}} \frac{\vert \Omega\vert ! \, (\vert \mathcal{L}^{p}\vert - \vert \Omega\vert - 1)!}{\vert\mathcal{L}^{p}\vert !} \left\{\Phi\left(\Omega\cup \xi\right)-\Phi\left(\Omega\right)\right\},
\end{equation}

where $\Phi\left(\Omega\cup \xi\right)-\Phi\left(\Omega\right)$ represents the marginal cost contribution of player $\xi$, and the weight $\frac{\vert \Omega\vert ! \, (\vert \mathcal{L}^{p}\vert - \vert \Omega\vert - 1)!}{\vert\mathcal{L}^{p}\vert !}$ represents the probability of occurrence of the coalition $\Omega$. Over all coalitions, the Shapley value gives the average marginal costs of a player $\xi$ over all coalitions $\Omega$ of congested lines.

The map $\Phi$ is given by the result of the optimal power flow computation. For each coalition $\omega$, the line limits are modified such that only the subset of lines in $\Omega$ leads to overloads. 
Given the cost of all coalitions, the pairwise difference of costs yields the marginal contributions of each player $\xi$ to that coalition.

The Shapley algorithm is depicted in Figure \ref{fig:flowchart}. 
First, we determine the overloaded lines by comparing the power flow results with the optimal power flows. 
Then, the costs of all coalitions of congested lines are computed with AC optimal power flow. 
Lastly, the Shapley values are computed, and potentially summed up to give the costs for each system operator.

\vspace{-1em}
\begin{center}
\tikzstyle{io} = [rectangle, rounded corners, minimum width=1cm, minimum height=0.5cm,text centered, draw=black]
\tikzstyle{startstop} = [trapezium, trapezium left angle=70, trapezium right angle=110, minimum width=2cm, minimum height=0.5cm, text centered, draw=black]
\tikzstyle{process} = [rectangle, minimum width=2cm, minimum height=0.5cm, text centered, draw=black]
\tikzstyle{decision} = [diamond, minimum width=2cm, minimum height=0.5cm, text centered, draw=black]
\tikzstyle{dummy} = [diamond, draw=white]

\tikzstyle{arrow} = [thick,->,>=stealth]

\begin{figure}[htbp!]
\centering
\scalebox{0.7}{
\begin{tikzpicture}[node distance=2cm]

    \node (input) [io] {Grid};
    \node (powerflow) [process, below of=input, yshift=1cm, text width=1cm] {PF};
    \node (OPF) [process, below of=powerflow, yshift=1cm, text width=1.2cm, fill=gray!15] {\acrshort{opf}};
    \node (overloads) [io, below of=OPF, yshift=1cm] {Set of congestions $\mathcal{L}^\textrm{cl}$};
    \node (out1) [process, right of=input, xshift=2.5cm, text width=3.5cm] {Operation costs $f(\Omega)$, \newline $\forall$ coalitions $\Omega\in\mathbb{P}(\mathcal{L}^\textrm{cl})$};
    \node (shap) [process, below of=out1, yshift=0.5cm, text width=3.5cm, fill=gray!15] {Compute shapley value $\Phi_\xi(f)\hspace{0.5em}$, $\forall$ lines $\xi\in\mathcal{L}^\textrm{cl}$};
    \node (out2) [io, below of=shap, yshift=0.5cm, text width=4cm] {Congested costs per congested line};
    
    \draw [arrow] (input) -- (powerflow);
    \draw [arrow] (powerflow) -- (OPF);
    \draw [arrow] (OPF) -- (overloads);
    \draw [arrow] (overloads) -- ++(2.0cm,0) |- (out1);
    \draw [arrow] (out1) -- (shap);
    \draw [arrow] (shap) -- (out2);


\end{tikzpicture}}
\vspace{1em}
\caption{The Shapley algorithm.}
\label{fig:flowchart}
\end{figure}
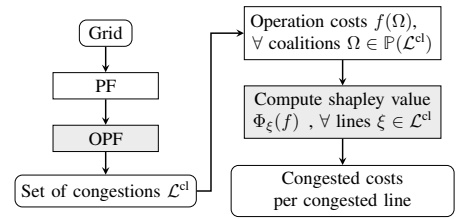
\end{center}
\vspace{-3em}

\subsection{AC OPF Formulation}
When calculating the costs $\Phi(\Omega)$ over all coalitions $\Omega\in\mathbb{P}(\mathcal{L}^{p})$, in each iteration an optimal power flow computation with modified line limits is performed. The line limits for all lines $(i,j)\in\tilde{\mathcal{L}} = \mathcal{L}\setminus\Omega^\textrm{c}$ are kept, while those for lines $(i,j)\in\Omega^\textrm{c}$, where $\Omega^\textrm{c} = \mathcal{L}^{p}\setminus\Omega$, are neglected. 
The DC OPF formulation can be found in \cite{voswinkel}.

We formulate the AC \acrshort{opf} problem following~\cite{Introduction} using polar coordinates, i.e.  $V_i=v_i e^{\textrm{j} \theta_i}$, where $v_i$ is voltage magnitude, $\theta_i$ is voltage angle, $p_i$ and $q_i$ denote active and reactive power at the bus $i$, while $p_{ij}$ and $q_{ij}$ denote active and reactive power flow along the line $(i,j)\in\mathcal{L}$. 
Hence, the resulting AC \acrshort{opf} problem can be written as
\begin{subequations}\label{eq::opf}
    \small
    \begin{flalign}
        \min_x \; & f(x) = \sum_{i \in \mathcal{N}}\left\{a_{i} \left(p^{g}_i\right)^2+b_{i} p^{g}_{i} + c_{i}\right\}&\label{eq::opf::obj}\\
        \textrm{s.t.}
        \; &p_i^g-p_i^l =v_i \sum_{k\in\mathcal{N}} v_k \left( G_{ik} \cos\theta_{ik} + B_{ik} \sin\theta_{ik} \right),\,&\forall i\in\mathcal{N}\label{eq::opf::pf::active}\\
        &q_i^g-q_i^l =v_i \sum_{k\in\mathcal{N}} v_k \left( G_{ik} \sin\theta_{ik} - B_{ij} \cos\theta_{ik} \right),\,&\forall i\in\mathcal{N}\label{eq::opf::pf::reactive}\\
        &p_{ij}= \; \ v_i^2G_{ij}-v_i v_j\left(G_{ij}\cos \theta_{ij}+B_{ij}\sin \theta_{ij}\right),&\forall (i, j)\in\tilde{{\mathcal{L}}}\label{eq::opf::line::active}\\
        &q_{ij}=-v_i^2G_{ij} - v_i v_j\left(G_{ij}\sin \theta_{ij}-B_{ij}\cos \theta_{ij}\right),&\forall (i, j)\in\tilde{{\mathcal{L}}}\label{eq::opf::line::reactive}\\
        & p_{ij}^2+q_{ij}^2\leq \overline{s}_{ij}^2,& \forall (i, j)\in\tilde{{\mathcal{L}}}\label{eq::opf::line::limit}\\
        &\underline u_i \leq u_i \leq \overline u_i,\;\;
        \underline p_i^g \leq p_i^g \leq \overline q_i^g,\;\;
        \underline q_i^g \leq q_i^g \leq \overline q_i^g,\;&\forall i\in\mathcal{N}\label{eq::opf::box}
\end{flalign}
\end{subequations}
with $x = (\theta,\,v, \,p^g,\, q^g)$, where $\theta_{ij}$  denotes the phase angle difference between buses $i$ and $j$, $p^g_i$ $q^g_i$ the power of generator at bus $i$, $p^l_i$ and $q^l_i$ the load at the bus $i$, and $G_{ij}$ and $B_{ij}$ are the real and reactive components of the bus admittance matrix. Therewith, \eqref{eq::opf::obj} denotes the quadratic function for the total generation cost, \eqref{eq::opf::pf::active}-\eqref{eq::opf::line::reactive} denotes the nodal power balance for each bus, \eqref{eq::opf::line::active}-\eqref{eq::opf::line::limit} represent the line limits w.r.t. the apparent power, and \eqref{eq::opf::box} puts upper and lower bounds on the state variables. 

\section{Experiments \& Results}
\label{sec:results}

We investigate how the DC and AC OPF influences the number of congestions, the redispatch costs, and the resulting Shapley values of grids with fixed loads. First, we show that AC OPF can lead to significant differences in the Shapley algorithm on small grids. Then, we apply the algorithm to large-scale power grids showing more effects.

The study uses grids of sizes 9 to 2383 buses, mostly from the PGLib library \cite{PGLib}. These grids are based on the IEEE test grids and have slightly modified parameters, including more realistic generation costs. 
For each case, we add several congestions, see Appendix \ref{sec:gridmodifications}, of which representative cases are portrayed in the present paper. 
The grid coordinates of large grids are constructed with the software yEd \cite{yEd} and do not represent real-world distances. 
Regarding the plots, generation is marked in green, loads in red, and the thickness of the blue lines indicates the amount of the power flows.

Computations are performed with the \matpower package in \matlab-R2022b \cite{MATLAB} on a standard desktop computer with \texttt{Intel\textsuperscript{\textregistered} 
i5-6600K CPU @ 3.50GHz} and 32.0 \textsc{GB} installed \textsc{ram}.

\subsection{DC vs. AC OPF in the Shapley algorithm}

\subsubsection*{Two effects of AC OPF on the Shapley values}
Our first major result is that AC OPF produces significantly different power flows compared to DC OPF, thereby altering the costs of each congestion, i.e. the Shapley value. Thus, AC OPF is not simply giving more exact solutions, but qualitatively different solutions than DC OPF.

\begin{figure}
    \centering
    \includegraphics[width=0.45\linewidth, trim = 1.3 1.3 1.3 1.3, clip]{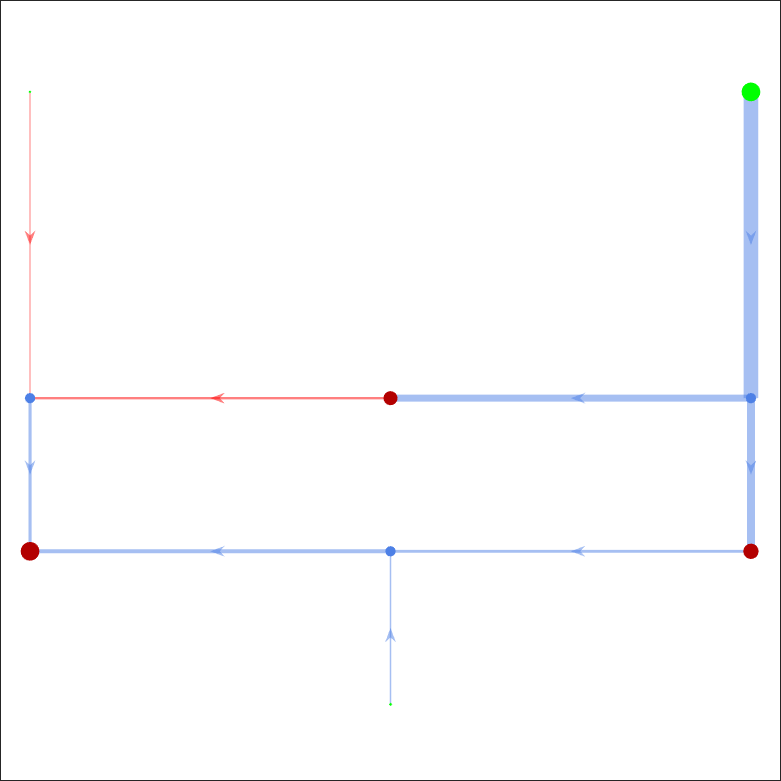}~
    \includegraphics[width=0.45\linewidth, trim = 1.3 1.3 1.3 1.3, clip]{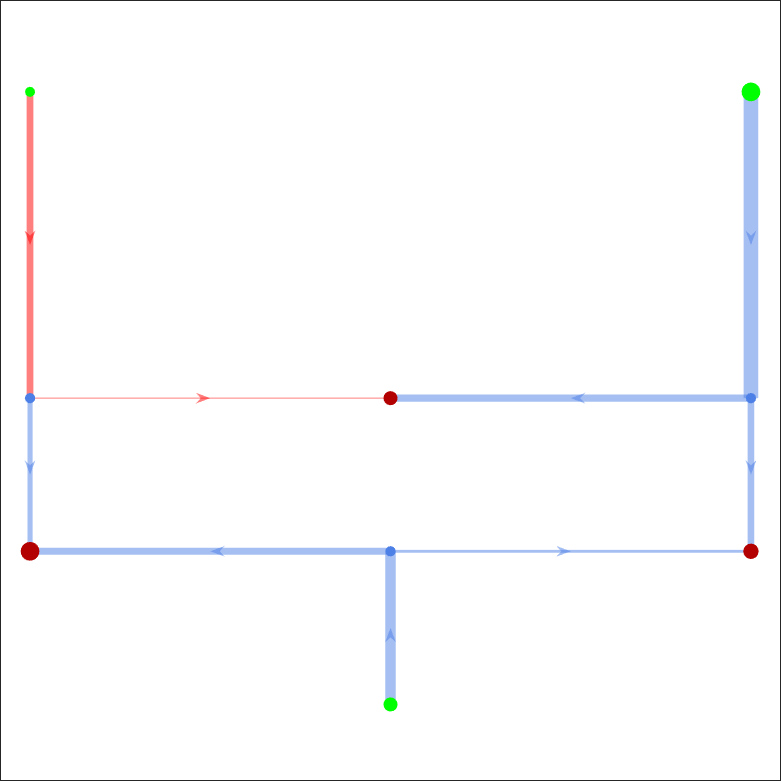}
    \caption{The images show the overloaded lines with DC OPF (left) and AC OPF (right).}
    \label{fig:case9_DCACdifference}
\end{figure}

A simple example of the 9-bus system is given in Figure \ref{fig:case9_DCACdifference}, the numerical values are given in Tables \ref{TB::numerical-result-DC} and \ref{TB::numerical-result-AC}. The grid is computed both with DC OPF (left) and AC OPF (right). Although the same lines $(1,2)$ are overloaded, their Shapley values are opposite; with DC OPF line $2$ is the main congestion, while AC OPF yields line $1$ as the main congestion. 
In more detail: According to the market demands and capacity of load and generation, represented by the power flow (PF) calculation, the generator at bus $1$ would produce $315$MW in the DC setting and $324$MW with AC OPF. Given the line limits of $(70 \text{MW},40 \text{MW})$, this yields overloads of $(245.0 \text{MW},100.2 \text{MW})$  and $(261.4 \text{MW},105.7 \text{MW})$, respectively. After redispatch, represented by the optimal power flow (OPF) computation including line limits, the power flows are $(13.1 \text{MW},40.0 \text{MW})$ and $(70.0 \text{MW},15.6 \text{MW})$, respectively. This yields the Shapley values $(0,123.5)$ (DC) and $(5.5,0)$ (AC). 

We can observe two effects. 
Firstly, the respective total redispatches differ significantly, and, against the expectation, the AC OPF has much lower costs. 
The second effect is that different lines are assigned Shapley values (costs) at all. While DC OPF yields that line $2$ should be enforced, AC OPF yields that line $1$ should be enforced. 
These effects are due to topological redistribution of the underlying graphs of the respective power grids. Hence, a simple extrapolation of DC OPF results might be insufficient or even counterproductive. Similar effects can be observed in larger grids. This leads us to the conclusion that AC OPF can have significantly different solutions of the Shapley value, and that decisions with effect in the real world should not be based on DC OPF.

\subsubsection*{Effects of AC OPF on large-scale power grids}

In large-scale power grids, AC OPF, too, defies the expectation that the values are simply more exact and that DC OPF is already a good approximation. Neither are the redispatch amounts always larger, nor are the congestions of the same amount and the same lines. 
Not even the runtimes are always higher for the more complex AC OPF, e.g. case793.

\begin{figure}
    \centering
    \includegraphics[width=0.45\linewidth, trim = 1.3 1.3 1.3 1.5, clip]{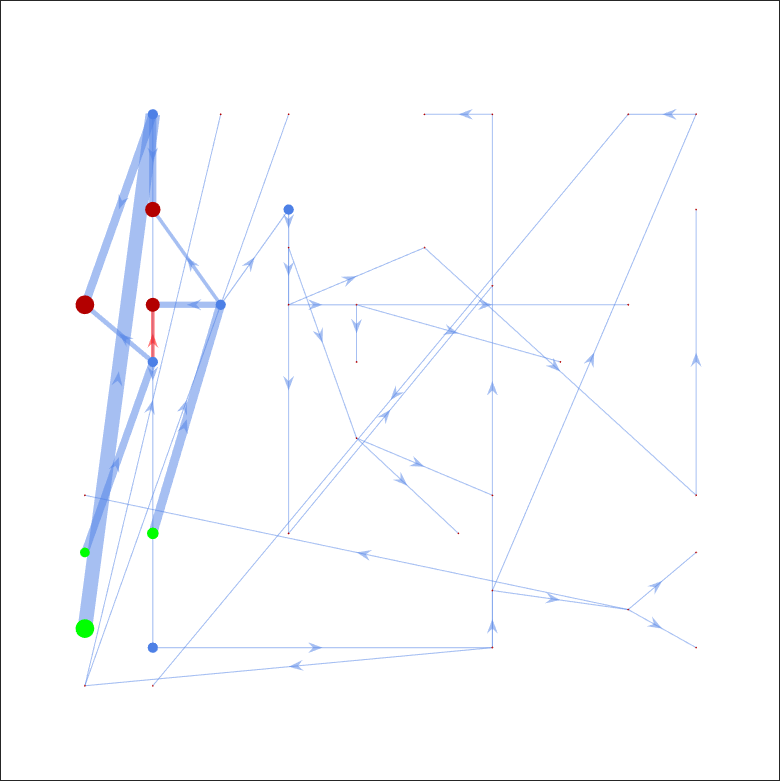}~
    \includegraphics[width=0.45\linewidth, trim = 1.3 1.3 1.3 1.5, clip]{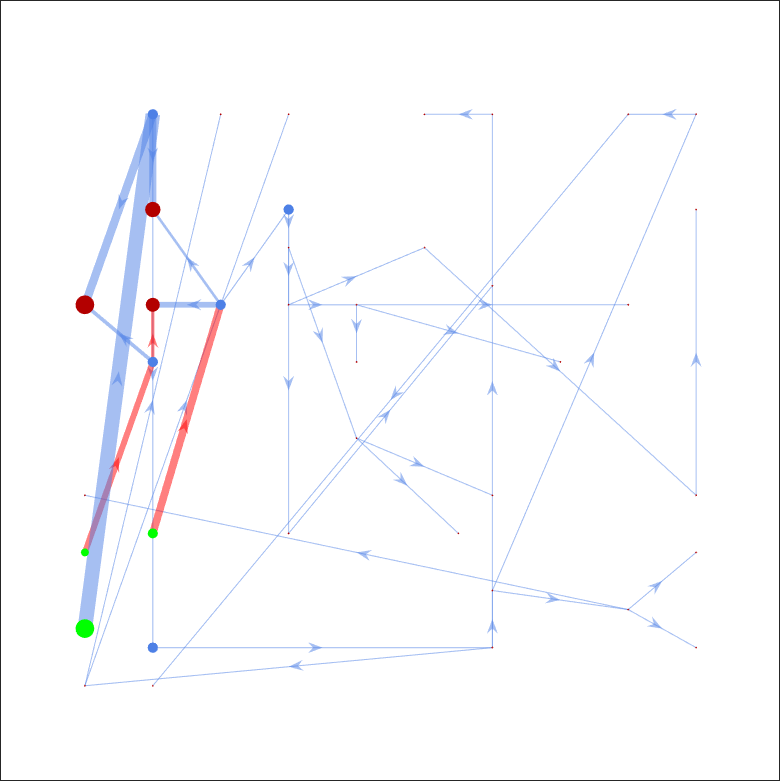}
    \caption{Case39 with congestions from DC and AC OPF.}
    \label{fig:case39}
\end{figure}

\begin{figure}
    \centering
    \includegraphics[width=0.45\linewidth, trim = 1.3 1.3 1.3 1.5, clip]{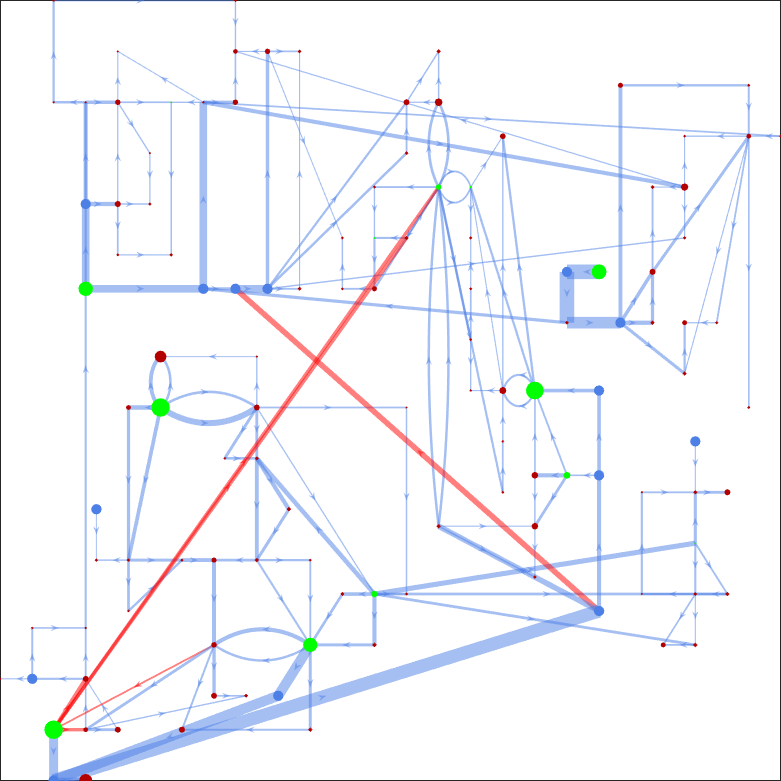}~
    \includegraphics[width=0.45\linewidth, trim = 1.3 1.3 1.3 1.5, clip]{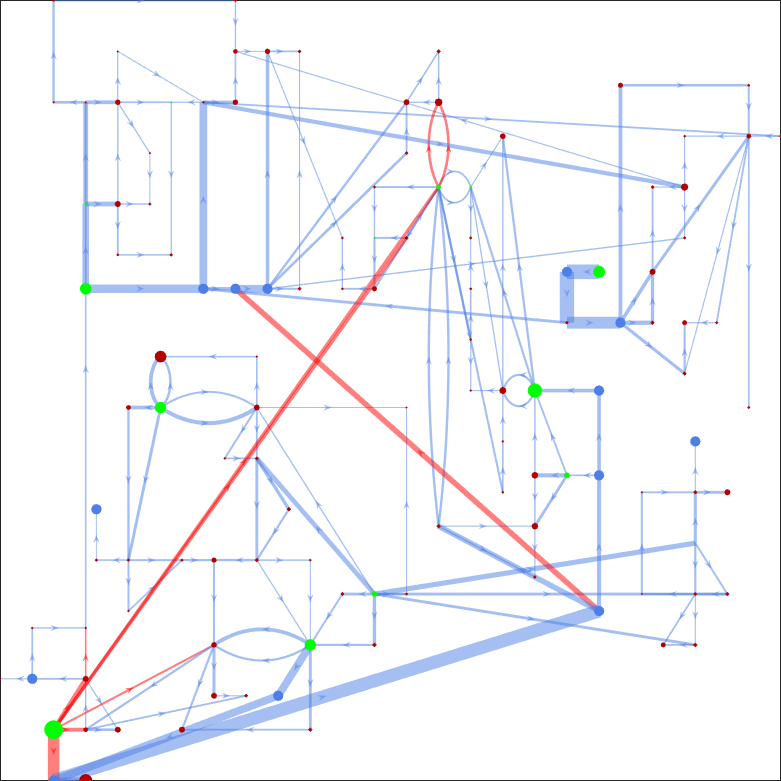}
    \caption{Case118 with congestions from DC and AC OPF.}
    \label{fig:case118}
\end{figure}

\begin{figure}
    \centering
    \includegraphics[width=0.45\linewidth, trim = 1.3 1.3 1.3 1.5, clip]{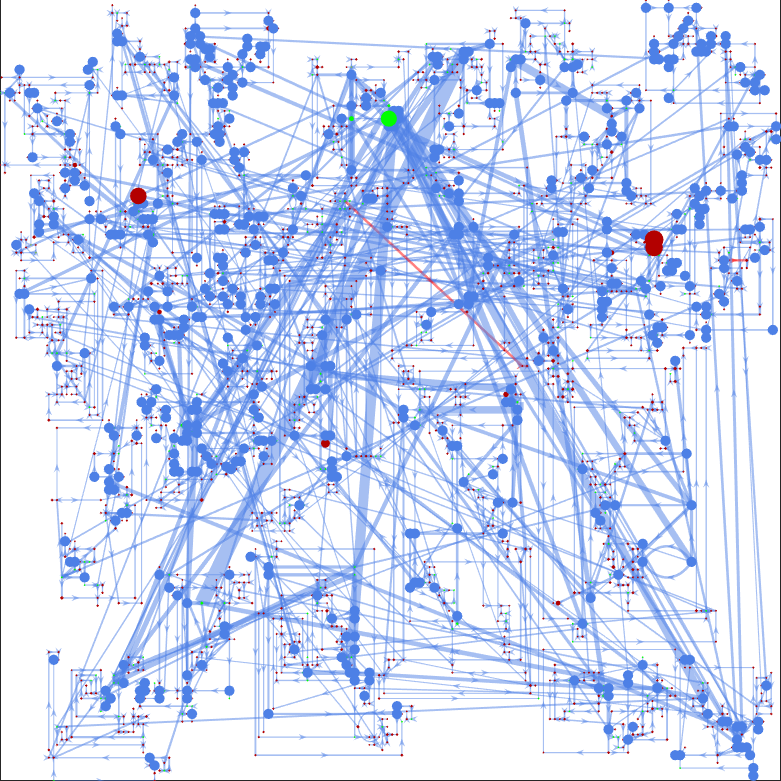}~
    \includegraphics[width=0.45\linewidth, trim = 1.3 1.3 1.3 1.5, clip]{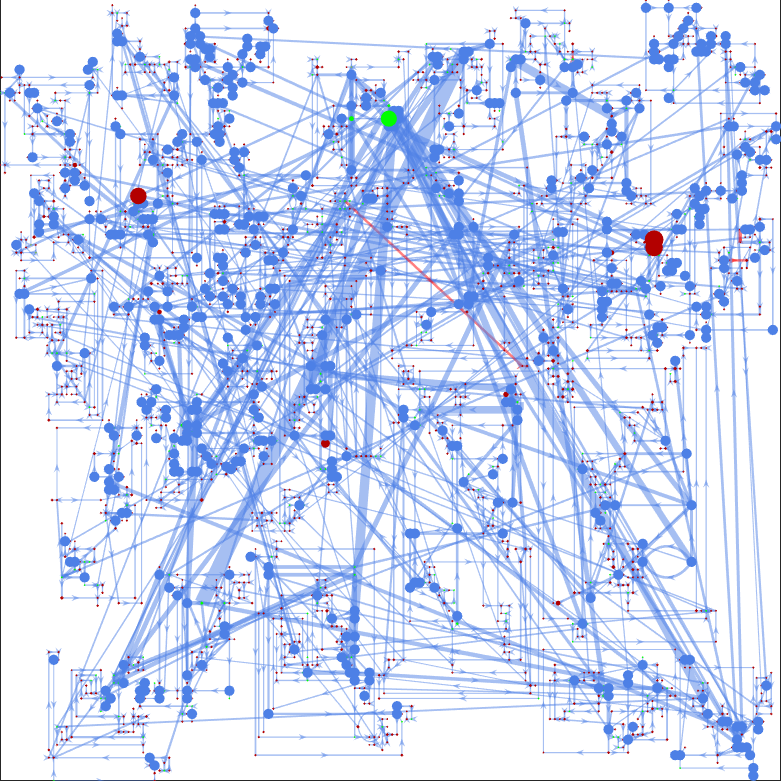}
    \caption{Case2383 with congestions from DC and AC OPF.}
    \label{fig:case2383_AC}
\end{figure}

Tables \ref{TB::numerical-result-DC} and \ref{TB::numerical-result-AC} show the simulation results for all cases with 9 to 2382 nodes (see Figures \ref{fig:case39}, \ref{fig:case118}, and \ref{fig:case2383_AC}). Some results are as expected; we often have higher overall system costs, higher redispatch costs and larger Shapley values. However, we observe three counter-intuitive effects.

Firstly, the number of congestions differs. As for most tested cases, AC OPF produces more congestions than DC OPF, see cases case39, case118, case1354, and case2383. However, case300 produces the same congestions with very similar values for both DC and AC OPF. 

Secondly, looking at case793, we see that AC OPF can also produce fewer congestions than DC, 3 compared to 7, and has lower redispatch costs, namely 0. The latter says that generation did not change when altering the line limits, compared to the market clearing. One explanation could be that AC OPF already outputs very different generation due to stricter constraints. However, the precise differences are up to grid topology.
Moreover, in case2383 we observe that, even though AC produces more congestions, the redispatch costs are still lower than with the DC approximation. 
Considering that, in the AI context, the Shapley value gives an interpretation of the importance of a feature, it can also give an interpretation of the relevance of a congested line in a power grid, given fixed loads and generation.

Thirdly, the Shapley values are not proportional to the overloads. We see that many Shapley values are $0$. This can be interpreted as follows: Given the market clearing, loads and generation capacities, the Shapley value assigns an importance to each line. It gives a hint to how critical the line is to changing the overall costs. 
Thus, it also says how much that line would have been influenced in the power flows of the original market clearing setting.

\begin{table*}[t]
    \caption{The Shapley values with the DC OPF approximation.} \label{TB::numerical-result-DC}
    \centering
    \scriptsize
    \renewcommand{\arraystretch}{1.5}
    \begin{tabular}{cc|c|c|c|c|c}\toprule
        Cases  & Lines & Congested Lines & Total system costs (\$) & Total redispatch costs (\$) & Shapley Values (\$) & Runtime (min) \\
        \hline
        case9  & 9 & (1,2) & 6566 & 123  & \textbf{(0,123)} & 0:00:56 \\
        case39 & 39 & (2) & 5334 & 3  & (3) & 0:00:49 \\
        case118 & 186 & (96,105,106,108,116,119) & 93132 & 42 & (0,0,42,0,0,0) & 00:16:36 \\
        case300 & 411 & (394,400) & 707405 & \textbf{1112} & (401,711) & 00:01:14 \\
        case793 & 913 & \textbf{(188,324,338,399,616,910,911)} & 253905 & 2529 & (0,0,0,0,2183, 330,16) & 0:26:00 \\
        case1354 & 1991 & (86,223,230,1067) & 1218096 & 2337 & (84,0,2253,0)& 00:01:42 \\
        case2383 & 2896 & (1816,2109,2110) & 1777927 & 1677.1 & (1662,15,0.1) & 00:01:42 \\
        \bottomrule
    \end{tabular}
\end{table*}

\begin{table*}[t]
    \caption{The Shapley values with the AC OPF.} \label{TB::numerical-result-AC}
    \centering
    \scriptsize
    \renewcommand{\arraystretch}{1.5}
    \begin{tabular}{cc|c|c|c|c|c}\toprule
        Cases & Lines & Congested Lines & Total system costs (\$) & Total redispatch costs (\$) & Shapley Values (\$) & Runtime (min) \\
        \hline
        case9 &  9 & (1,2) & 7748 &  5 &  \textbf{(5,0)} & 0:01:25 \\
        case39 & 39 & (1,2,4) & 5666 & 274 &  (2,231,41) & 0:11:00 \\
        case118 & 186 & (66,67,96,105,106,107,108,109,116,119) & 97213 & 236 & (0,0,0,40,196,0,0,0,0,0) & 2:50:51 \\
        case300 & 411 & (394,400) & 720612 & \textbf{886} & (336,550) & 00:05:36 \\
        case793 & 913 & \textbf{(324,399,616)} & 255547 & \textbf{0} &  \textbf{(0,0,0)} &  0:12:00 \\
        case1354 & 1991 &  (86,223,230,299,1067,1868) & 1258843 & 2776 & (11,0,1521,1244,0,0) &  2:32:20 \\
        case2383 & 2896 & (305,309,322,1382,1816,2109,2110,2862) & 1859054 & 515 & (0,0,250,0,104,161,0,0) & 15:16:43 \\
        \bottomrule
    \end{tabular}
\end{table*}

\subsection{Discussion}

Further insights on the variability of the Shapley value in the algorithm are:

\subsubsection*{What influence does the initial power flow setting have on the Shapley values?}
The power flow 
determines where the overloads will be as the power flow results are compared against the optimal power flow results. The market clearing should be reasonably realistic, i.e. close to the OPF solution.

\subsubsection*{Can the Braess' Paradox cause negative Shapley values?}
No. The Braess' Paradox can only appear in AC settings where either the reactance $X$ is modified, or a new line is added \cite{Braess}. 
Increasing or decreasing the line capacity yields a real extension or contraction of the feasibility set.

\subsubsection*{What do zero Shapley values mean?} \label{sec:SVzero}
A Shapley value of $0$ tells us that the redispatch does not cause any difference in generation costs, solely in the power flows through the grid. Hence, there is no real redispatch of generation. 
There is one exception when several generators have the exact same generation costs, however that hardly ever occurs. 

\subsubsection*{Is the feed-in priority of RES considered?}
Yes. It can be achieved by setting the costs of RES lower than those of conventional generators.

\subsubsection*{Why are Shapley values disproportional to their overloads?}
The Shapley values do not represent the proportional amount of overload on that line, they represent whether the congestion on that line changes the generation schedule, and, by that, cause large redispatch. Some lines are more relevant in the given setting than others. In that sense, the Shapley value offers an interpretation on the impact of lines.
\section{Conclusion}
\label{sec:conclusion}

Redispatch costs rise, and the question of their equitable distribution and exact computation become more and more relevant. 
Concerning fair distribution, the Shapley value has been proposed and applied to small concatenated power grids. However, all current methods for redispatch use the DC approximation of optimal power flow in this context.
In this study, we ask what influence DC versus AC optimal power flow has on the Shapley values, i.e. the costs of each congestion, and apply the approach on large-scale transmission systems.
The results show that using AC OPF significantly changes the results, i.e. the congestions, the redispatch costs and the Shapley values.
Notably, AC OPF does not in all cases lead to more congestions, higher redispatch costs and higher Shapley values. These effects are presumably due to complex grid topologies and lead to the realization that DC OPF cannot simply be extrapolated to real AC grids, but needs to be handled with care.
Furthermore, we give some insights into the variability of the Shapley values.
We conclude with the recommendation of using AC optimal power flow in future research and industry application at the TSO.

Further research could include a real-world case study, for example with the transmission grid of Germany. However, since real grids are very large and may lead to convergency issue, future work could investigate how the grids can be reduced such that AC OPF converges. Moreover, the effects of AC OPF could differ for meshed grids versus concatenated grids, e.g. transmission and distribution grids. Further, using the former distributed AC OPF approach, the effect of different partition sizes could be investigated. Lastly, flexibilities such as battery storage could prevent congestions and reduce redispatch costs, and add to AC convergence.

\section{Appendix}\label{sec:appendix}

\subsection{Grid modifications} \label{sec:gridmodifications}

For reproducibility purposes, we list the modifications that we added to the grids to create congestions:
For the IEEE case9 we set the generation to buses $1,2$ and $3$ and the load to buses $4,5,6,7,8,9$. Generation costs are linear with coefficients $(30,0), (25,0), (20,0)$. The branch limits of lines $1,2$ are $70,40$. 
For the IEEE case39 we added at lines $1$ and $4$ the branch limits $100$ and $90$.
For the IEEE case300 we set the branch limits to 1000. 
The PGLib case118 is unchanged. 
In the PGLib case793 we increased the line limits by a factor of $1.4$ to ensure feasibility.
The PGLib case1354 remains unchanged.
In the IEEE case2383 we added at lines $24,169,292,321,322$ and $1381$ the limits $100,200,200,100,10$ and $10$ to ensure feasibility.




\section*{Acknowledgments}
The authors thank our German local TSO for providing the research funds and questions on real-world power grids.



\end{document}